\documentclass[sigconf,nonacm]{acmart}

\usepackage{todonotes}
\usepackage{caption}
\usepackage{subcaption}

\newcommand{\mpara}[1]{\medskip\noindent{\bf #1}}

\newcommand{\model}{$\Phi$}
\newcommand{\query}{$q$}
\newcommand{\doc}{$d$}
\newcommand{\xdoc}{$\hat{d}$}

\setcopyright{acmcopyright}
\copyrightyear{2018}
\acmYear{2018}
\acmDOI{10.1145/1122445.1122456}

\acmConference[Woodstock '18]{Woodstock '18: ACM Symposium on Neural
  Gaze Detection}{June 03--05, 2018}{Woodstock, NY}
\acmBooktitle{Woodstock '18: ACM Symposium on Neural Gaze Detection,
  June 03--05, 2018, Woodstock, NY}
\acmPrice{15.00}
\acmISBN{978-1-4503-XXXX-X/18/06}

\date{}
\begin{document}

    \title{BERT Rankers are Brittle: a Study using Adversarial Document Perturbations}

\author{Yumeng Wang}
\email{wang@l3s.de}
\affiliation{%
  \institution{L3S Research Center}
  \city{Hannover}
  \country{Germany}
}

\author{Lijun Lyu}
\affiliation{%
   \institution{L3S Research Center}
   \city{Hannover}
   \country{Germany}} 
\email{lyu@l3s.de}

\author{Avishek Anand}
\affiliation{%
   \institution{L3S Research Center}
   \city{Hannover}
   \country{Germany}} 
\email{anand@l3s.de}

\renewcommand{\shortauthors}{Wang and Lyu, et al.}

\begin{abstract}

Contextual ranking models based on BERT are now well established for a wide range of passage and document ranking tasks.
However, the robustness of BERT-based ranking models  under adversarial inputs is under-explored.
In this paper, we argue that BERT-rankers are not immune to adversarial attacks targeting retrieved documents given a query.
Firstly, we propose algorithms for adversarial perturbation of both highly relevant and non-relevant documents using gradient-based optimization methods.
The aim of our algorithms is to add/replace a small number of tokens to a highly relevant or non-relevant document to cause a large rank demotion or promotion.
Our experiments show that a small number of tokens can already result in a large change in the rank of a document. 
Moreover, we find that BERT-rankers heavily rely on the document start/head for relevance prediction, making the initial part of the document more susceptible to adversarial attacks.  
More interestingly, we find a small set of recurring adversarial words that when added to documents result in successful rank demotion/promotion of any relevant/non-relevant document respectively.
Finally, our adversarial tokens also show particular topic preferences within and across datasets, exposing potential biases from BERT pre-training or downstream datasets. 
\end{abstract}


\maketitle

\begin{figure}[t!]
      \centering
      \includegraphics[width=1.0\linewidth]{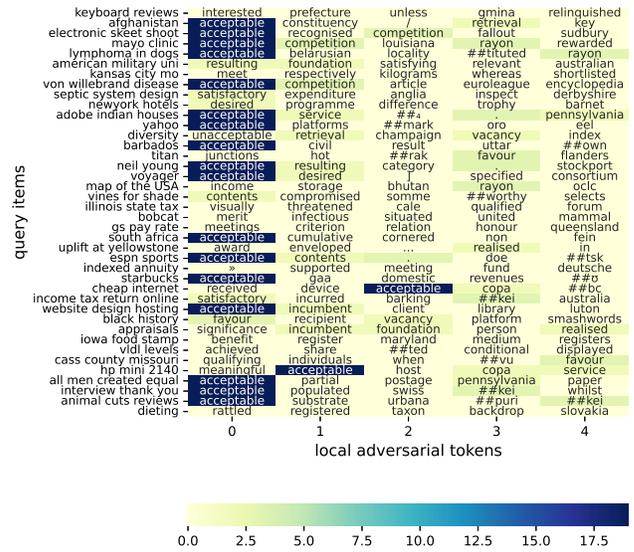}
      \caption{The five adversarial tokens added to the beginning of the highest-ranked document for 40 queries from ClueWeb09, selected by local ranking attack method to demote the document. Specific tokens frequently recur across queries. The frequency is denoted by the color.}
      \label{fig:heatmap}
    \end{figure}

\section{Introduction}
\label{sec:intro}

Adversarial examples are small deliberate perturbations to an input instance that can lead to wrong predictions.
There is a growing amount of work that has shown that over-parameterized neural models can easily be fooled/attacked for a variety of machine learning tasks~\cite{kurakin2016adversarial,tramer2017ensemble,miyato2016adversarial}.
Adversarial examples not only showcase the limitations of the underlying trained model by exposing non-intuitive and unreliable results, but they also expose the potential biases of the model or training corpus. 
 
The ability to generate adversarial examples for ranking models is of substantial interest to search engines and e-commerce websites, that are increasingly using neural ranking models.
Additionally search engine optimization (SEO) companies benefit from adversarial terms that can potentially improve the ranking of any arbitrary document.
Finally, adversarial examples also provide an insight into the inner workings of the models in terms of spurious correlations resulting from pre-training and fine-tuning procedures.

So far, however, the question of adversarial perturbations for text ranking models has not been addressed in detail.
Existing works on document perturbations for text ranking deal predominantly with black-box attacks with limited applicability~\cite{raval2020one},  human-assisted adversarial examples~\cite{goren2018ranking} or for interpretability of rankers~\cite{singh2020model}.
In this work, we propose white-box adversarial attacks on BERT-based rankers by perturbing text documents.

Unlike adversarial perturbations on images where arbitrarily small changes are possible in the image space, text data is different and arguably more challenging.
Due to the discrete nature of language, small changes in the input space in the form of word additions or replacements can easily cause a big difference in embedding vectors. 
In this paper, we define the problem of adversarial ranking attacks that generate adversarial terms/tokens that when added to a retrieved document greatly shift the rank of the document.
Note that we choose to perturb long documents since it is hard to detect instead of perturbing queries.

We consider two attack scenarios -- (a) where a highly relevant document is \textit{demoted}, and (b) where a lowly ranked document is \textit{promoted}. 
Additionally, we define the scope of adversarial ranking attacks on a per-query level and on an entire query workload.
Corresponding to these scopes we design a \emph{local ranking attack} and a \textit{global ranking attack} that aim to maximize the rank shifts.
Our document perturbation approaches are adapted from gradient-based token search algorithms for local~\cite{ebrahimi2017hotflip} and the global~\cite{wallace2019universal} attack. Accessing the parameters of a real-life ranking model is hard. Instead, we choose the BERT-style ranker as the victim, because it's been intensively applied in many text-related tasks including ranking for its superior success. Such big gain of performance usually comes at the cost of vulnerability against adversarial attack, as well as the lack of interpretability.  Figure~\ref{fig:heatmap} showcases the five adversarial tokens selected by the \textit{local ranking attack} to demote the documents for $40$ anecdotal queries.  More details can be found in Section~\ref{sec:token-analysis}. 



 We conduct extensive experiments where we attack the BERT ranker on ClueWeb09 and TREC-DL datasets. 
 Our results show that both local and global attacks can cause significant rank shifts, exposing the fragility of BERT rankers with the addition of as few as five tokens. 
 Our further post-hoc analysis suggests the adversarial tokens from local ranking attack recur across different queries, sometimes transcending across datasets. 
 Finally, we find a majority of adversarial tokens fall into a group of sensitive topics like \textit{ethnics}, \textit{diseases} or \textit{natural disasters}, indicating potential dataset biases. The source code is publicly available~\footnote{\url{https://github.com/menauwy/brittlebert}}.  

\section{Related Work}
\mpara{Adversarial Attacks in NLP}. 

There are two major lines of attacking approaches in NLP tasks such as text classification and natural language inference. The first is without knowing model architectures and parameters, namely under the black-box setup, the attacker uses predefined heuristic rules to generate \textit{natural} (in human perspective) substitutes of words or sentences, so that to fool the victim model. The prevalent heuristics include synonym replacement~\cite{jin2019bert, ren2019generating}, mask-and-fill by contextualized language models~\cite{ li2020contextualized, li2020bert}, human-in-the-loop strategy~\cite{wallace2019trick}, etc. On the other hand, white-box attack assumes full access to model parameters. Thus the attacker can use gradient signals to guide the searching process within a few rounds of forward and backward computation~\cite{ebrahimi2017hotflip, XU2020103641}. Due to the discrete nature of human languages, gradient-based approaches are efficient yet lack of fluency compared to rule-based methods. For more related adversarial attacking methods, we point the reader to recent surveys~\cite{zhang2020adversarial, huq2020adversarial}.

\mpara{Universal Triggers}. Unlike above methods, \citet{wallace2019universal} generate input-agnostic universal triggers using HotFlip~\cite{ebrahimi2017hotflip} for NLP models and datasets. Triggers are tokens or phrases, injecting such trigger to any input text can mislead the victim model to a target prediction. For high-dimensional neural networks, it's unsurprising to find tokens from a large vocabulary that always result in a particular prediction. Moreover, \citet{wallace2021poisoning} also argue it's possible to plant any trigger by poisoning the training data with a small set of crafted instances containing such trigger. After retraining, the model will pick up the shortcuts induced by the trigger and make the same prediction whenever the trigger occurs in the input. We are more interested in this universal trigger because it exposes higher potential risks if a single phrase can not only impact the model prediction, but also apply to the whole dataset.

\mpara{Adversarial Attacks in Text Ranking.} There has been limited research regarding adversarial attacks on ranking models, especially text ranking models. 
Closest to our work is~\citet{raval2020one}, who propose a model-agnostic document-perturbation procedure for rank demotions.
Our work is different in that (i) we use gradient-based method to search candidate tokens to replace and be replaced; (ii) we consider more scenarios such as document demotion and promotion for transformer models and (iii) more importantly, we focus on local as well as global adversarial tokens as an attempt to discover potential dataset and model biases.  

\section{Adversarial Attacks on Neural Rankers}
\label{sec:ranking-attacks}

We operate on the common retrieve and re-rank framework for document ranking and focus on adversarial perturbations in the re-ranking phase.
Our aim is to generate adversarial examples for already trained neural ranking models by perturbing retrieved documents at user-specified positions in the document.
We consider two attack scenarios: (1)~\textbf{rank demotion} for high-ranked documents (e.g., ranked in top 50, since it makes no sense to demote the lowest-ranked document), and (2)~\textbf{rank promotion} for low-ranked documents (e.g., the last 50 positions). 
The scope of an adversarial ranking attack can be either local (on a given query) or global (on an entire query workload).
Towards this, we firstly envision the \textbf{local ranking attack}  where we intend to perturb a retrieved document given a query.
Secondly, we formulate the \textbf{global ranking attack} where we intend to generate adversarial tokens for an entire workload of queries.
We describe our approach to realize both attacks in detail in the following subsections.

\subsection{Local Ranking Attack} 

Given a query \query{} and a text input \doc{} from dataset $\mathcal{D} = \{Q, D\}$ where $q\in Q$ and $d\in D$, the ranking model \model{} computes the relevance score of \doc{} w.r.t \query{} by \model{}$(q, d)$ and the ranking position $\Pi_{d}$ is obtained by comparing all relevance scores of a list of pre-retrieved documents. 
We intend to craft an adversarial example \xdoc{} for \doc{}, by adding $i$ tokens $\mathbf{x} = \{x_1, x_2, \cdots, x_i\}$ or replacing $i$ tokens with $\mathbf{x}$ in $d$, so that the rank difference $|\Pi_{d}-\Pi_{\hat{d}}|$ is maximized. We initialize with $i$ placeholder tokens (i.e. [MASK]) if the adversarial tokens are to be added and the position of perturbation is user-defined. 
We denote such perturbation by $\hat{d} = d \odot \mathbf{x}$. Specifically,  we minimize the relevance score to demote $d$  (or  maximize it to promote $d$):
\begin{align}
\label{eq:attack-instance}
  \underset{\mathbf{x}}{ \text{arg\, min}} \ \Phi (q, d\odot \mathbf{x})
\end{align}
To ensure that number of document perturbations is small, we choose  $i \leq 20$ in our experiments. 
To solve the above objective function, we adapt gradient-based token search algorithms like HotFlip~\cite{ebrahimi2017hotflip} to find the adversarial tokens. 
More concretely, we compute the gradient of the relevance function and use the gradient signals to search for the tokens over the entire vocabulary $\mathcal{V}$. 
Due to the discrete nature of tokens, we approximate the relevance changes induced by tokens from the whole vocabulary $\mathcal{V}$ using the first-order Taylor expansion as following: 

\begin{align}
    \label{eq:attack-tokens}
  \underset{v\in \mathcal{V}}{ \text{arg\, min}} \ [v - x_i]^\top \nabla_{x_i} \Phi (q, d\odot x_i)
\end{align}

Note that $v - x_i$ is element-wise subtraction on the embedding dimension. 
To search for multiple adversarial tokens, we then apply beam search to extract the top $i$ candidate tokens similarly as~\cite{wallace2019universal}. 
The best $\mathbf{x}$ is obtained by repeating equation~\eqref{eq:attack-tokens} until the relevance score decreases no more.  

\subsection{Global Ranking Attack} 

Rather than generating adversarial tokens for a particular query-document instance, in global ranking attacks, we aim to find tokens $\mathbf{x}$ that are adversarial to the entire query set. 
In other words, adding such tokens can promote or demote the respective highly relevant or non-relevant document for any query in the dataset. 
Towards this, we minimize the expected relevance score for all queries in the demotion scenario as:
\begin{align}
    \label{eq:attack-universe}
  \underset{\mathbf{x}}{ \text{arg\, min}} \ \mathbb{E}_{(q, d) \sim \mathcal{D}} [ \Phi (q, d\odot \mathbf{x})]
\end{align}

The global adversarial tokens are selected by using the same gradient-based search strategy as equation~\eqref{eq:attack-tokens} and updating on all queries until the average relevance score decreases no more.

We choose to ensure imperceptible perturbation by the number of adversarial tokens. This sort of adversarial texts are inevitably not as natural as those generated by black-box methods, or with synonyms constrains. We take the trade-off because attacking is not the only purpose of the work, but also discovering the spurious recurrences of adversarial tokens.

\section{Experiments}


\subsection{Experimental Setup}
 
\mpara{Datasets and Model} This paper focuses on the widely successful BERT-based rankers for experiments. 
We fine-tune the BERT (bert-base-uncased) model on (i) ClueWeb09 and (ii) MSMARCO passage ranking datasets using a pairwise training loss. Specifically, we construct the input consists of a query and a document separated by the [SEP] token, with which the BERT ranker then predicts a relevance score. The ranker is trained to maximize the margin between the scores of a relevant and non-relevant input pair. We execute our local and global attacks on a query workload of 200 real-world  ClueWeb09 queries. 
For MSMARCO, we randomly selected 1000 queries from the development set and the 200 queries from the TREC-DL test set (all denoted as TREC-DL for simplicity) to attack. For each query, we consider the top 50 as relevant and the rest as non-relevant, for each group we randomly sample 10 documents to commit rank demotion and promotion respectively. Thus, we deal with a retrieval depth of $100$ for all our experiments.

\mpara{Metric.} We measure the absolute rank shifts normalized by the maximum  shift distance (e.g., a document at rank 10 can only be demoted maximally 90 positions) as $\mathtt{NRS} = \frac{|\Pi_d - \Pi_{\hat{d}}|}{|\Pi| - \Pi_d}$ where $|\Pi|$ is the retrieval depth (i.e. $|\Pi| = 100$). 


\subsection{Adversary Attack Effectiveness}

\begin{figure}[t!]
    \centering
    \includegraphics[width=0.8\linewidth]{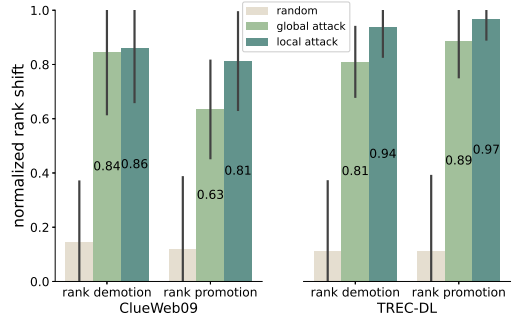}
    \caption{Attack effectiveness. Local vs. Global vs. Random}
    \label{fig:general}
\end{figure}

\begin{figure}
    \centering
    \includegraphics[width=0.8\linewidth]{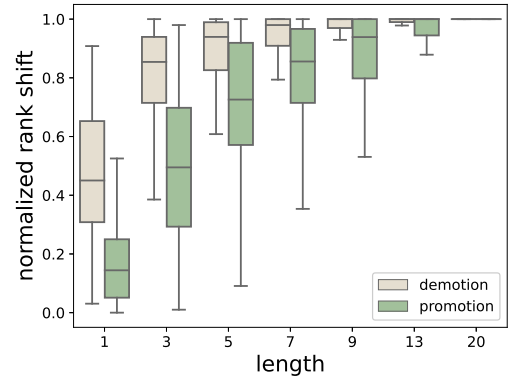}
    \caption{Impact of adversarial token lengths}
    \label{fig:length}
\end{figure}

\begin{figure}
    \centering
    \includegraphics[width=0.8\linewidth]{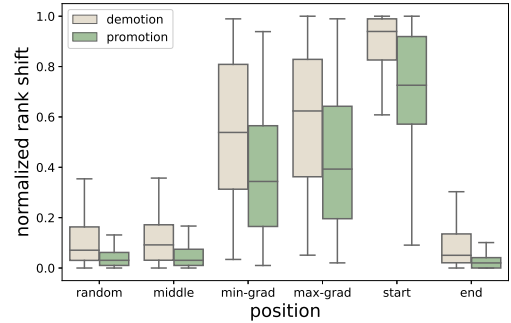}
    \caption{Impact of adversarial token positions}
    \label{fig:position}
\end{figure}


In this experiment, we measure the rank shifts of the relevant and non-relevant documents when a fixed small number of tokens ($5$ tokens for ClueWeb09 and $3$ for TREC-DL) are added to the document's start (refer to Figure~\ref{fig:general}). 
We compare our two proposed approaches to a baseline where the same number of randomly selected tokens are used as adversarial tokens. 
As Figure~\ref{fig:general} shows, for both datasets even a small number of tokens can cause significant rank shifts in comparison to the random baseline. 
We observe that TREC-DL shows significantly more pronounced rank shifts when compared to ClueWeb09. 
This observation can be attributed to the fact that TREC-DL contains shorter passages than ClueWeb09, and thus it is more sensitive to input perturbations. 
Specifically in TREC-DL, carefully selecting three words is sufficient (on average) to promote irrelevant passages that are ranked below 50-th position into the top-3. 
Similarly, selecting three words can demote the high-ranked passages to the bottom-six ranks (rank 94 -- 100). 
Meanwhile, in ClueWeb09, adding five tokens to the retrieved documents results in relatively smaller yet significant rank shifts of around $80$. 
It is also clear that the local ranking attack outperforms the global attack method. 
This observation is perhaps not surprising since the global adversarial tokens are updated based on the entire query workload, trading-off effectiveness for better generalization.


\subsubsection{Effect of the number of adversarial tokens}
 
A logical follow-up question is how many tokens are needed to cause a significant rank shift. 
Towards this, we conduct an experiment where we vary the number of allowable tokens added ( 1 -- 20 tokens) to the start of the retrieved documents and measure the average rank shifts. 
In the following, whenever we mention rank shifts, we mean \emph{average} rank shifts for ease of exposition.
In Figure~\ref{fig:length}, we present the rank shifts for ClueWeb09. We omit TREC-DL due to space constraints.

The results indicate that as few as seven tokens are sufficient to cause a rank shift of above $90\%$.
Also, increasing the number of allowable tokens monotonously increases rank shifts with noticeable marginal returns after $5$ tokens.
More stark is the observation that $20$ tokens are sufficient to cause nearly the maximum rank shift.  
Note that for long ClueWeb09 documents (that typically contain more than 512-token BERT input limit), even a single token is sufficient to cause a rank shift of $46$.
Finally, we note that rank shifts in both directions are equally affected, indicating the fragility of BERT ranking models to minor document perturbations. 

\mpara{Insight 1:} \textit{We observe that both datasets show significant rank shifts by adding a small number ($\leq 5$) of tokens. }

\subsubsection{Effect of Token Positions}

Until now, we have focused on adding the tokens to the start of the document. 
We also conducted an ablation study for different attack positions within the document. 
We considered the following attack positions -- (i) the \textbf{start}, (ii) the \textbf{end}, the positions where the original tokens have the  (iii) \textbf{highest} and (iv) the \textbf{lowest} $i$ ($i$=5) gradient scores.
We also looked at (v) \textbf{random}  positions and (vi) approximately the \textbf{middle} position in the document (computed from the text length divided by $2$).
Note that all the attack positions assume that the document fits in the 512 token-limit of the BERT input. 
In case document is longer than the input limit, only the truncated head of the document is considered in our experiments, a common experimental design choice for BERT-based rankers. 
The results for ClueWeb09 are reported in Figure~\ref{fig:position}, which indicates a significant sensitivity of ranking models to document perturbations on the start position. 
Namely, modifying the start of the texts causes much higher rank shifts than other positions. 
We hypothesize that BERT rankers might automatically associate higher relevance to the terms present at the beginning of the document.
We leave the detailed analysis of this claim to future work.

\mpara{Insight 2:} \textit{We observe a significant sensitivity of ranking models to document perturbations on the start position.}

\subsection{Adversarial Token Analysis}
\label{sec:token-analysis}

Since the local attack approach generates $i$ adversarial tokens for each query independently, we performed a post-hoc analysis of the local adversarial tokens to check for potential recurrence patterns.
We analyzed both rank demotions and promotions with an expectation that seemingly disparate queries should have low overlap in adversarial tokens.

\begin{figure}[t!]
      \centering
      \includegraphics[width=1.0\linewidth]{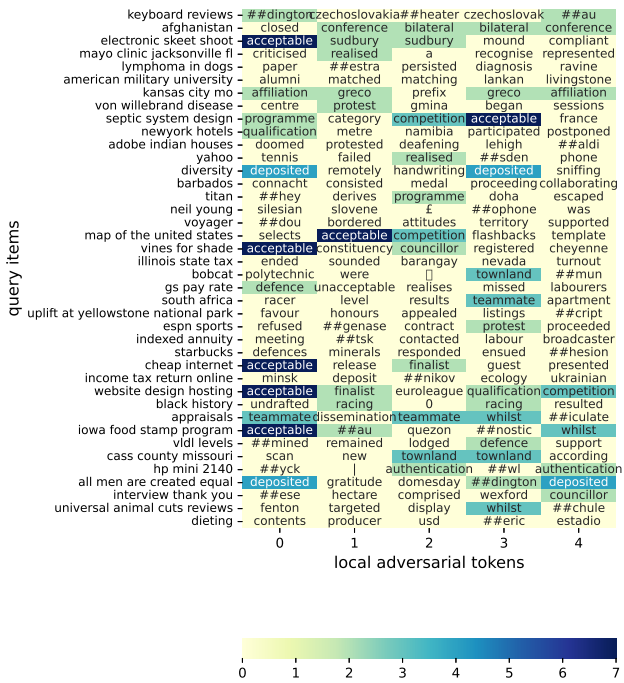}
      \caption{The five adversarial tokens replacing the tokens in documents at \textit{max-grad} positions for 40 queries from ClueWeb09, selected by local ranking attack method to demote the document. Specific tokens frequently recur across queries. The frequency is denoted by the color.}
      \label{fig:heatmap_inplace}
    \end{figure}

\mpara{Rank Demotion.} We randomly sample $40$ queries from ClueWeb09 and list the $5$ tokens selected to demote the top-ranked documents in Figure~\ref{fig:heatmap} (due to space limitations, we omit the rest of the queries). 
Surprisingly, we observe some particular tokens recur with a high frequency, such as the term ``acceptable". 
In other words, the ranking model associates the term ``acceptable" with a negative relevance signal irrespective of the query in the dataset. To avoid the position bias, we also show the results selected by the same algorithm to replace tokens with the highest gradient scores for the same documents in Figure~\ref{fig:heatmap_inplace}. It suggests such negative relevance of particular terms also hold irrespective of the attacking position. 
On the other hand, we also find the term ``acceptable" in the five global adversarial tokens for the same dataset.
This suggests that both local as well as global ranking attack are capable of uncovering recurring terms like  ``acceptable", ``competition," and ``rayon", that negatively impact the relevance score of a document.
Interestingly, our experiments also show that adding $5$ of the most frequent tokens to all top-ranked documents causes, on average of a rank drop of $83$ positions, compared to a rank drop of $79$ positions by global adversarial tokens.   

\mpara{Rank Promotion.} Unlike demoting relevant documents, adversarial tokens to promote partially relevant documents are expected to contain a high fraction of query terms.
This is both understandable, expected, and in fact, serves as a sanity check for our attack algorithms.
Unsurprisingly, we find that around half of the adversarial tokens are indeed query tokens. 
However, similar to rank demotions, we find that there are also frequent repetitions of particular tokens for rank promotions. 
Specifically, natural-disaster-relevant tokens such as ``tornadoes" and ``hurricane" are the most favored by the model for many different queries. Note that the $5$ tokens selected by the global attack, are ``hindusim", ``earthquakes", ``childbirth", ``tornadoes" and ``Wikipedia", showing high similarity to the frequent local adversarial tokens.

This recurrence of potentially non-relevant tokens is an important application of our study and reflects the potential biases of the BERT-ranking model.
On further investigation, we find that these terms are not frequent in the corpus and do not have a high mutual information with the relevance label.  
The reason for the recurrence of the term is out of the scope of this work and is left for future investigations.


\mpara{Insight 3:} \textit{We observe that local adversarial tokens recur across multiple queries and also in across datasets.}

\subsection{Pre-Training Bias vs Dataset Bias} 
\label{sec:dataset-bias}

\begin{figure}[t!]
      \centering
      \includegraphics[width=0.9\linewidth]{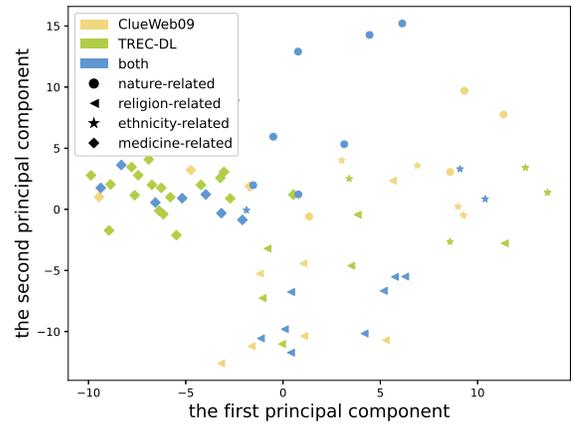}
      \caption{Frequently appeared adversarial tokens from local ranking attack for both datasets. }
      \label{fig:pca}
    \end{figure}

Generalizing from the token level, we also noticed the existence of additional recurrent topics from the local adversarial tokens.
In Figure~\ref{fig:pca}, We present the adversarial tokens added to the document start in a 2D visualization (using PCA) for both datasets. 
We omit rank demotion since the tokens do not show prominent topical patterns, and we consider only tokens with a minimum support of $2$.
Figure~\ref{fig:pca} indicates some tokens related to \textit{nature}, \textit{religion}, \textit{ethnicity} and \textit{medicine} are chose for both datasets. We omit presenting the selected tokens from diverse positions in addition to the document start since they show a similar topic preference as Figure~\ref{fig:pca}.
Since we use the same BERT model and fine-tune it on two different downstream datasets, we hypothesize that the model might exhibit some bias due to the pretraining process. 
On the other hand, it could also be possible that the datasets retain some topic preferences. 
Figure~\ref{fig:pca} shows that TREC-DL has a strong preference of \textit{ medicine} while for ClueWeb09, \textit{religion} and \textit{nature} are slightly more dominant. Via manual observation of the $200$ queries from ClueWeb09, we found out $8$ queries relevant to \textit{nature}, $14$ related to \textit{ethnicity}, the queries about diseases are as many as around $24$. Additionally, some sporadic queries about cities or hotels resulted in many documents relevant to \textit{nature} and \textit{religion}. All of the relevant queries and documents caused the frequent occurrences of the topic specific adversarial tokens.    For TREC-DL, we leave a more automatic method for topic extraction to future work due to the large training corpus.       

\section{Conclusion}
In this paper, we adapt the gradient-based adversarial attack algorithm on BERT ranking models to deliberately demote and promote documents in per-query level and the entire query workload. Our experimental results suggest a minor perturbation on the text documents can shift the rank by a large margin, exposing the fragility of BERT ranking models. Moreover, we also show the BERT model is particularly sensitive to the perturbations on the document start. Finally, we conducted a post-hoc statistical analysis on the adversarial tokens and found out a set of specific tokens recurring across queries and datasets. Our findings implicate the potential biases of BERT pretraining and downstream datasets. We hope our study can raise more awareness of the risk of applying large pretrained models for ranking task.  For future work, we plan to include more diverse models and datasets, endeavoring to uncover biases for large corpus and neural ranking models.   






\begin{acks}
This work is partially supported by DFG Project AN 996/1-1.
\end{acks}

\bibliographystyle{ACM-Reference-Format}
\bibliography{sigir}

\appendix

\end{document}